\documentclass[aps,twocolumn,showpacs,superscriptaddress]{revtex4}
\usepackage{dcolumn}
\usepackage{graphicx}
\usepackage{verbatim} 
\usepackage{color}
\usepackage{ulem}
\usepackage{amsmath}
\usepackage{soul}

 \usepackage[utf8]{inputenc}

\newcommand{\editor}[2]{%
  \expandafter\newcommand\csname #1note\endcsname[1]{%
    \textcolor{#2}{(\textbf{#1:} ##1)}}%
  \expandafter\newcommand\csname #1\endcsname[1]{%
    \textcolor{#2}{##1}}%
  \expandafter\newcommand\csname #1cancel\endcsname[1]{%
    \textcolor{#2}{\sout{##1}}}%
  \expandafter\newcommand\csname #1change\endcsname[2]{%
    \textcolor{#2}{\sout{##1} ##2}}%
  \newenvironment{#1text}{\color{#2}}{\color{black}}
}
\definecolor{BluBondi}{rgb}{0.00,0.58,0.71}
\editor{CC}{BluBondi}

\begin{document}

\title{Strong magnetic proximity effect in Van der Waals heterostructures driven by direct hybridization}
\author{C. Cardoso}
\affiliation{S3 Centre, Istituto Nanoscienze, CNR, Via Campi /a, Modena (Italy)}
\author{A. T. Costa}
\affiliation{International Iberian Nanotechnology Laboratory, 4715-330 Braga, Portugal}

\author{A. H. MacDonald}
\affiliation{Physics Department, University of Texas at Austin, Austin, Texas 78712, USA}
\author{J. Fern\'andez-Rossier}
\altaffiliation{On leave from Departamento de F\'{\i}sica Aplicada, Universidad de Alicante, 03690,  Sant Vicent del Raspeig, Spain }
\affiliation{International Iberian Nanotechnology Laboratory, 4715-330 Braga, Portugal}

\begin{abstract}
We propose a new class of magnetic proximity effects based  on the spin dependent hybridization between
the electronic states at the Fermi energy in a non-magnetic conductor and 
the narrow  spin split bands of a ferromagnetic insulator.  Unlike conventional exchange proximity, we show this hybridization proximity effect has a very strong influence  on the non-magnetic layer and can be further modulated by application of an electric field.  
We use DFT calculations to illustrate this effect in graphene placed next to
a monolayer of CrI$_3$, a  ferromagnetic insulator. We find strong hybridization of the graphene bands with 
the narrow conduction band of CrI$_3$ in one spin channel only. We show that our results are
robust with respect to  lattice mismatch and twist angle variations. 
Furthermore, we show that an out-of-plane electric field 
can be used to modulate the hybridization strength, paving the way for applications.
\end{abstract}

\maketitle


Proximity effects in hybrid systems transfer an electronic property, intrinsic to a certain material, to an adjacent material in
which that property is absent. For example, the surface of  a material that otherwise has no electronic order can become superconducting, or magnetic, when placed in contact with superconductors or a ferromagnet. 
Proximity effects are an important resource in the design of  quantum matter in artificial heterostructures. For instance, topological superconductivity can be induced by taking advantage of either the superconducting proximity effect in semiconducting nanowires~\cite{lutchyn2010} or magnetic proximity effects on a superconductor surface~\cite{nadj2014}.  

With the advent of two-dimensional (2D) materials~\cite{geim2007,castellanos2016,frisenda2018}, proximity effects have 
become particularly relevant since they can alter the electronic properties of an entire  crystal.
The discovery of ferromagnetic 2D crystals~\cite{Huang:2017aa}, such as CrI$_3$, and the demonstration of Van der Waals heterostructures that combine them with non-magnetic 2D crystals~\cite{Klein18,Huang:2017aa}, motivate the present work.  
Specifically, there are now several experimental papers exploring the 
spin proximity effects induced by Van der Waals ferromagnetic insulators on non-magnetic 2D crystals, including semiconductors~\cite{zhong2017,norden2019,lyons2020}, graphene~\cite{tang2020,ghiasi2020,wu2021,lyu2022} and superconductors~\cite{,kezilebieke2020,kang2021}, as well as other exciting possibilities, such as 2D ferromagnetic metals in contact with antiferromagnetic insulators~\cite{zhang2020}, and Kitaev  materials in contact with graphene~\cite{mashhadi2019}.

The ferromagnetic proximity effect is most often modeled by adding spin-splitting to the bands of the non-magnetic material
to account for exchange interaction with a ferromagnet. In van der Waals materials 
density functional theory (DFT) calculations predict values of the proximity exchange splitting as large as 70~meV~\cite{qiao2014}, 
whereas the values reported experimentally are 
%
%
rather small, with induced spin splittings typically in the range of few meV.
These yield interesting but modest changes in the 
optical~\cite{zhong2017,norden2019} and transport~\cite{tang2020,ghiasi2020} properties of the non-magnetic material.  
Here we propose a complementary type of spin proximity effect, based on spin-dependent hybridization of the 
bands of a conducting material with those of a spin-split insulating ferromagnet. 
This type of proximity effect is always relevant when charge transfer occurs between magnetic and 
non-magnetic layers, and we show that it can have a radical impact in the properties of the non-magnetic material.
Importantly, it can be tuned electrically, opening the door to new device concepts. 

The proposed hybridization proximity effect arises from the spin dependent coupling 
between the atomic orbitals that form the conduction band of the conducting material and a set of  
localized spin split states in the ferromagnet.  In many cases it will 
significantly alter the energy bands and the mobility of only one of the spin channels of the conductor. 
Hybridization proximity occurs when a spin-split band of the ferromagnetic insulator 
is aligned with the Fermi energy of the conductor but, as we discuss below, this resonant condition 
occurs naturally by self-alignment of bands in a wide class of magnetic insulators.  

\begin{figure}
\centering

\includegraphics[clip,width=0.48\textwidth]{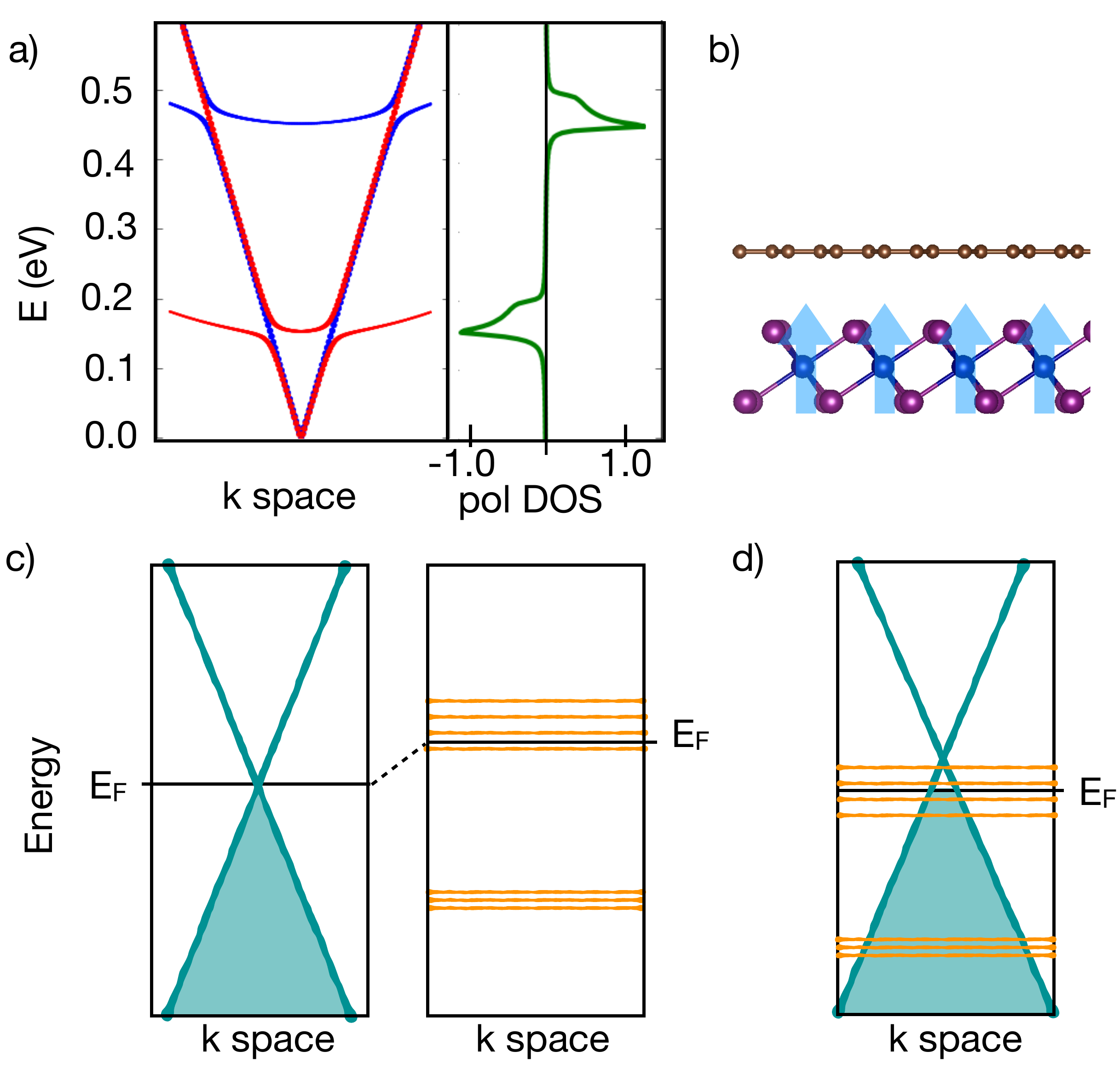}
\caption{\label{scheme}
(a) Left panel: energy bands of a toy model of a 2D conductor on top of magnetic insulator. 
Red and blue symbols indicate spin. Right panel: 
 $(\rho_{\uparrow}-\rho_{\downarrow})/(\rho_{\uparrow}+\rho_{\downarrow})$
 where $\rho_{\sigma}$ is the spin dependent density of states projected over the 2D conductor. The spin polarization in the non conductor is dramatically larger in the narrow energy windows where degeneracy with the insulating bands leads to anti-crossings;  (b) Graphene 
 on CrI$_3$ as an example of a 2D conductor on top of magnetic insulator (c) Schematic band structures of a 2D conductor (light green) and an insulator (orange), in the region around the Fermi level; (d) Self-alignment mechanism of the bands when proximity between
 the 2D conductor and the insulator is accompanied by charge transfer.} 
\end{figure}

To be explicit we consider in the following the case of the two dimensional conductor graphene, 
in proximity with a ferromagnetic insulator, like CrI$_3$ (see Figure~\ref{scheme}). 
The essence of the proposed effect is independent of the dimensionality of both the magnetic and the non magnetic material,
but the consequences will be more dramatic when the non-magnetic materials is two dimensional, 
and the magnetic insulator is sufficiently thin to allow integration in a dual gate structure with top and bottom gate electrodes.

The  microscopic principle of the hybridization proximity effect can be discussed in terms of a two band 
model Hamiltonian which neglects spin-orbit coupling.  The spin $\sigma=\uparrow,\downarrow$ Hamiltonian
\begin{equation}
{\cal H}=\left(
\begin{array}{cc}
\epsilon(\vec{k}) & V_{\vec{k}} \\
V_{\vec{k}}^*  &  E_\sigma(\vec{k})
\end{array}
\right)    
\end{equation}
where $\epsilon(\vec{k})$ is the spin-independent band dispersion of the non-magnetic conductor,
$E_{\sigma}(\vec{k})$ is the spin dependent energy band of the insulating ferromagnet and 
$V_{\vec{k}}$ is interlayer hybridization between these two bands.
In ferromagnetic insulators both the conduction and the valence band can have 
either majority or minority spin.  Interlayer coupling leads to two types of effects in the non-magnetic material. 
First,  away from the narrow energy window where the bands of the ferromagnetic material lie,  
hybridization leads to a perturbatively small 
spin-splitting  given by 
\begin{equation}
\delta(\vec{k})=|V_{\vec{k}}|
^2\left(\frac{1}{\epsilon(\vec{k})-E_{\uparrow}}-\frac{1}{\epsilon(\vec{k})-E_{\downarrow}}\right)
\label{splitting}
\end{equation}
Second,  in the energy window where the bare bands of both materials are degenerate, 
there is strong hybridization that opens up a gap.    
Since $E_{\uparrow}(\vec{k})\neq E_{\downarrow}(\vec{k})$, this gap only occurs in one spin channel, at a given energy. Thus,  if the Fermi energy lies in one of these resonance windows, the spin polarization of the conductor is effectively 100\%.

We now make two crucial observations.  First,  charge transfer occurs at the interface of the 2D conductor and the magnetic insulator 
when the difference in work functions is larger than the insulator's gap. 
As a result, either the valence or the conduction band of the magnetic insulator has to {\it self-align} with the Fermi energy of the conductor.  For this reason the hybridization proximity effect is relevant in a wide class of 
magnetic insulators, with spin polarized valence or conduction band. This is the central point of this work.  
Second, exact alignment between bands of the conductor and the magnetic insulator can be further tuned by 
application of an electric field perpendicular to the interface, which can result in
a strong change in the strength of the hybridization proximity effect and therefore
in a modulation of the spin splitting (Eq. \ref{splitting}). 

The electrostatics of the interface are roughly described by the Wigner-Bardeen rule~\cite{wigner1935}
which asserts that electrons will be transferred from the material with the smallest work function
$W_1$ to the material with the largest work function $W_2$.   
 When the difference in work function is larger than the energy gap,
we can estimate the charge transfer using a parallel plate estimation of the potential drop, 
in which extra electrons and extra holes are localized in two parallel planes separated at a distance $d$. 
The  areal density of excess carriers per layer, $\delta n$ is then related to the work function difference by 
\begin{equation}
W_2 - W_1 = e^2 \frac{\delta n}{\epsilon_0} d  \label{eq:1}
\end{equation}
where $e$ is the electron charge and $\epsilon_0$ is the vacuum dielectric constant.  Importantly, a finite $\delta n$ implies
charge injection into the ferromagnetic insulator and, thereby, alignment of its bands with the Fermi energy of the 
2D conductor.  In the following we substantiate this estimate with first principles DFT 
calculations for a hetero-bilayer formed by a ferromagnetic insulator, CrI$_3$, and graphene. 
This system has been studied experimentally~\cite{tseng22}.  For the sake of computational simplicity we choose a monolayer of CrI$_3$, but our results would hold for graphene on top of bulk CrI$_3$, given both the local nature of hybridization 
proximity, which involves predominantly the interaction between adjacent layers. 

CrI$_3$ is a ferromagnetic insulator with a  majority spin conduction band~\cite{soriano2020}. 
The Cr atoms are embedded in iodine octahedral cages that split the Cr $d$ levels into a $t_{2g}$ triplet and an $e_g$ doublet. 
The Cr oxidation state is +3, so that $S=3/2$ Cr$^{+3}$ ions have 3 electrons in the spin majority $t_{2g}$ orbitals  
that are strongly hybridized with the iodine $p$ orbitals to form the valence band~\cite{Lado_2017}. 
The conduction band of CrI$_3$ is formed by the spin majority $e_g$ bands, 
resulting in relatively non-dispersive bands with a large spin-splitting produced by intra-atomic exchange. 
The work functions of graphene and CrI$_3$ are $ W_G= 4.6$~eV~\cite{Filleter_2008, Datta_2009, Shi_2009} and $W_{\rm CrI_3} = 5.8$~eV 
respectively. Therefore, it is expected that electron transfer occurs from graphene to the conduction band of CrI$_3$~\cite{zhang2018}, self-aligning the narrow spin-split conduction band of CrI$_3$ and the Dirac cones of graphene.

In order to describe the hybridization proximity effect more quantitatively, 
we now carry out density functional theory simulations using the plane-wave and pseudo-potential implementation provided by the Quantum-ESPRESSO package~\cite{Giannozzi2009,Giannozzi2017}. We employed projector augmented wave (PAW) pseudo-potentials, and 
the Perdew-Burke-Ernzerhoff (PBE)~\cite{Perdew_96} exchange-correlation functional with Grimme-D2~\cite{Grimme_06} van der Waals corrections.

\begin{figure}[t]
\centering
\includegraphics[clip,width=0.45\textwidth]{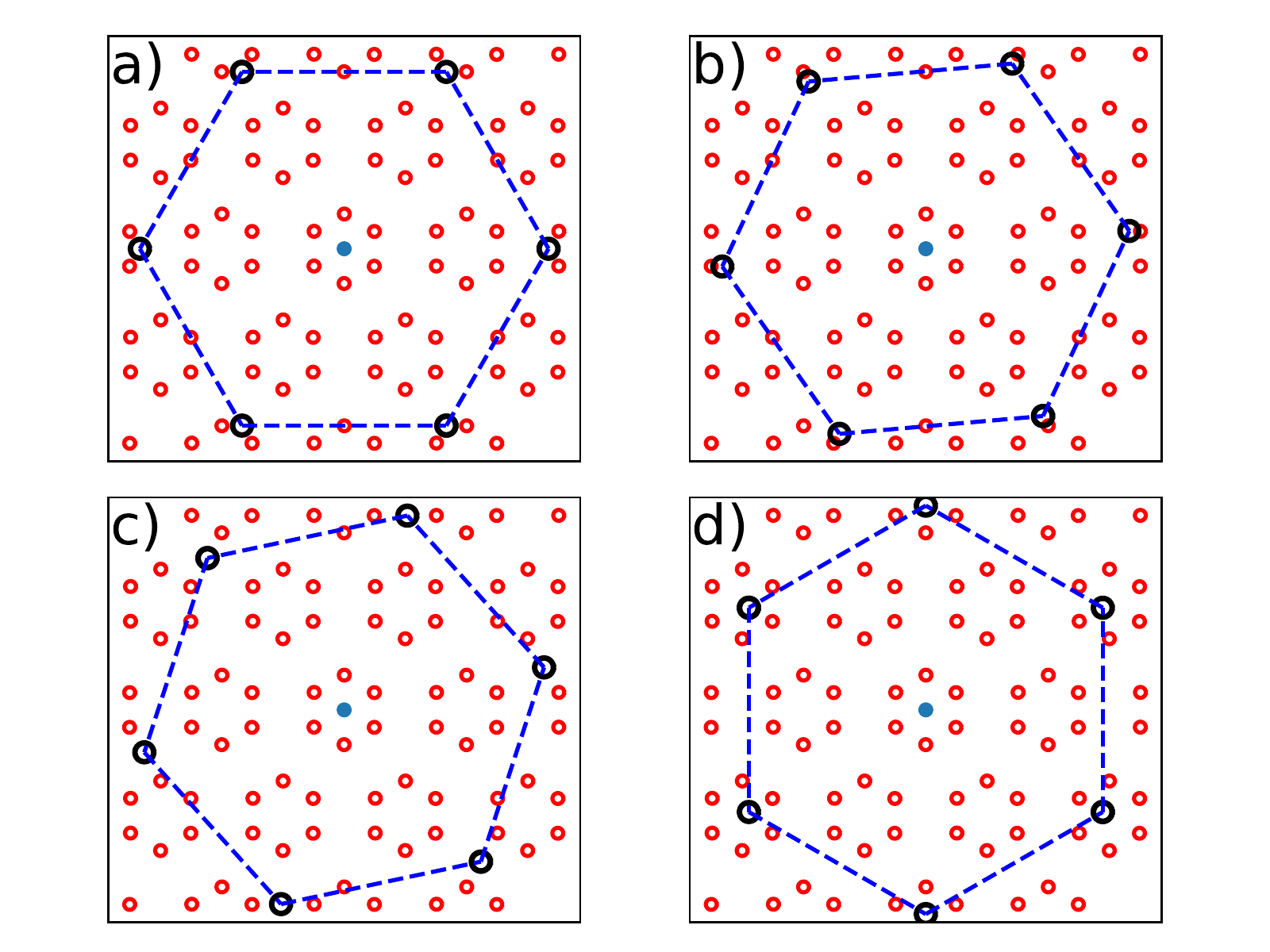}
\caption{\label{BZ}
Extended-zone representation of the Fermi surfaces (FS) of CrI$_3$ (red) and graphene (black), assuming charge transfer but ignoring interlayer hybridization. Each panel corresponds to a different rotation angle between graphene and CrI$_3$: a) $90^\circ$, b) $95^\circ$, c) $102^\circ$ and d) zero. The blue dot marks the position of the $\Gamma$ point. The dashed line marks the boundary of the first BZ of graphene. The lattice parameters used for these plots were $a_C=2.4$~\AA\ and $a_{Cr}=7.0$~\AA.}
\end{figure}

We first neglect inter-layer hybridization by solving Kohn-Sham equations for isolated CrI$_3$ and graphene layers, but
taking account of the expected electron density transfer $\delta n=1.9 \times 10^{13}\mathrm{cm}^{-2}$ from graphene to CrI$_3$.
Graphene therefore has hole pockets centered on the $K$ and $K'$ points in momentum space that enclose 
approximately 1\% of its Brillouin zone.  We find that CrI$_3$ has six electron pockets located away from the 
high-symmetry points of the Brillouin zone (BZ) that cover approximately 8\% of its much smaller Brillouin-zone area.  
Strong interlayer hybridization occurs when the Fermi surfaces of the 
two layers, smeared in energy by $\sim |V_{\vec{k}}|$, intersect~\cite{bistritzer2011,Shi23}.  
In Fig.~\ref{BZ} we plot the graphene and CrI$_3$ Fermi surfaces 
reduced to the first BZ of graphene, for different twist angles. It is apparent that, for most angles, the graphene Fermi surface pockets are very close or right on top some Cr$I_3$ pockets. 
Therefore, we  expect a moderate dependence of the interlayer hybridization on twist angle in this system.  

\begin{figure}
\centering
\includegraphics[clip,width=0.48\textwidth]{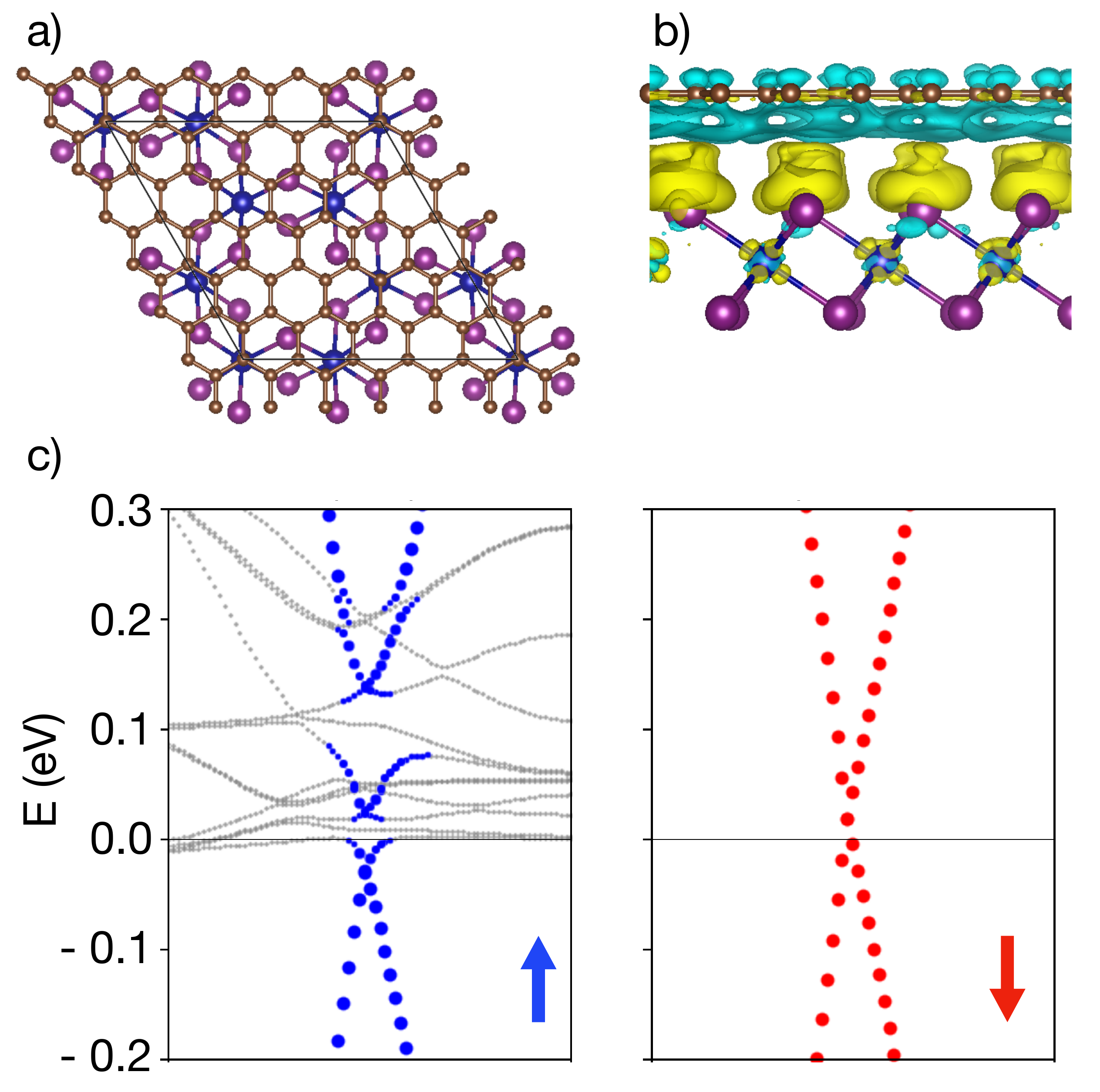}
\caption{\label{bands}
(a) Top and (b) side view of the Graphene/CrI$_3$ structure. In the side view we show the charge density difference between graphene/CrI$_3$ and isolated graphene and CrI$_3$ layers. (c) Energy bands of Graphene/CrI$_3$ monolayer for the region around the Dirac point. The blue (spin up) and red (spin down) dots correspond to the bands projected on the atomic orbitals of C, with the size of the dots depending on the magnitude of the projection.
}
\end{figure}

In order to support the above qualitative analysis for incommensurate CrI$_3$/graphene bilayers, we carried
out a DFT calculation for the coupled bilayer with a low-order commensurate unit cell.  
We consider a 5$\times$5 graphene supercell with 
$a_C=2.4572$~\AA\ (instead of  $2.4612$~\AA\ obtained for isolated graphene)
on top of a suitably oriented $\sqrt{3} \times \sqrt{3}$ CrI$_3$ supercell with $a_{Cr}=7.095$~\AA\ (instead of $7.008$~\AA).
The resulting energy bands for the CrI$_3$/graphene VdW structure are shown in Fig.~\ref{bands}c). 
As expected from the argument above, the spin minority  channel (in red) shows only the pristine graphene Dirac cone, since
the  spin majority states of CrI$_3$ are far away in energy.  
In contrast, in the spin majority channel (in blue) in the graphene cone is {\it strongly} hybridized with the CrI$_3$ conduction band 
shown in grey.  In this situation, we expect a strong spin dependence of the electronic transport:
whereas spin-minority channel has a pristine Dirac cone, the graphene states of the spin majority channel are scrambled with the narrow conduction band of CrI$_3$.  Given the Mott insulating nature of CrI$_3$~\cite{mcguire2015}, conduction in the spin majority channel is very likely strongly suppressed, compared with the spin minority graphene Dirac cone.  
This dramatically different landscape for the two spin channels at the Fermi energy 
constitutes the essence of the hybridization spin-proximity effect.

We now turn our attention to how to electrically control the band alignment between the non-magnetic conductor and the magnetic insulator. The band alignment is controlled by the magnitude of the charge transfer. 
The vertical shift of the bands at both sides, estimated with a jellium model in the Hartree approximation, would be given by $\delta V= (2\pi e\delta n d)(\epsilon_0)$, where $\delta n$ is the areal density of transferred electrons which
has opposite signs for opposite layers.
So, if we apply an out-of-plane electric field to compensate for the electrochemical charge transfer, this will reduce $\delta n$, and thereby will push the (hole rich) graphene layer down in energy and the (electron rich) magnetic insulator layer up in energy. Application of an electric field, for example with a single back-gate, would also inject additional carriers into the graphene/magnetic insulator bilayer.  This can be avoided using a dual gate device~\cite{zhang2009,taychat2010} which 
permits one to independently control the carrier injection and the voltage drop at the bilayer.

Our DFT  calculations with no applied electric field show that charges accumulate in the carbon atoms and in the iodine atoms of the top layer.  From DFT we also obtain an average distance between C and I of  d=3.54~\AA. 
Using Eq. \ref{eq:1} we  estimate the areal density of electron that transfer from graphene to CrI$_3$, 
 $\delta n=1.9 \times 10^{13}cm^{-2}$.  From DFT, we obtain a value of the integrated charge difference of 
 0.9$\times 10^{13}$~cm$^{-2}$, not far from the simple model estimate. 
 
\begin{figure}[t]
\centering
\includegraphics[clip,width=0.48\textwidth]{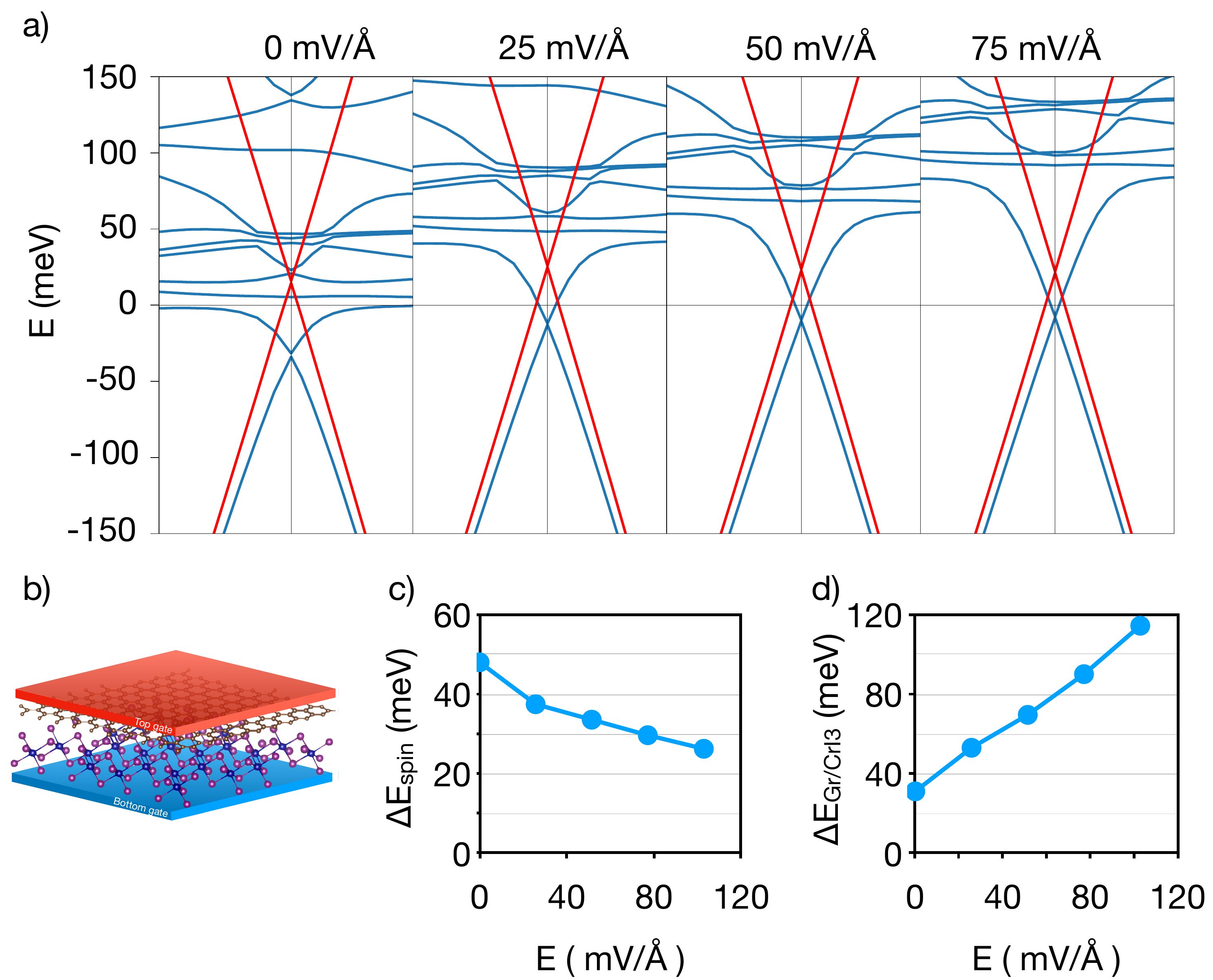}
\caption{\label{bands_zoom}
(a) Band structure computed for Gr/CrI$_3$ with increasing applied field;
(b) Schematic illustration of dual gate control of Gr/CrI$_3$;
(c) Energy difference between the two spin channels at K;
(d) Energy difference between the average position of the Dirac point and the bottom of the first conduction band.
}
\end{figure}

We have also carried out DFT calculations with an out-of-plane electric field. 
Since the charge in the system remains constant, our calculations describe 
an experimental situation where top and bottom gate voltages are chosen so 
as to make the carrier injection zero~\cite{lei2021}. The evolution of the energy bands for the Gr/CrI$_3$ heterobilayer as a function of electric field  are shown in Fig.~\ref{bands_zoom}. We consider fields up to E=100~meV/\AA, close to the limit
that can be achieved without reaching dielectric breakdown. 
The polarity of the field is chosen to reduce the charge transfer that occurs spontaneously at the Gr/CrI$_3$ interface. Therefore the displacement field moves electrons back from the conduction band of CrI$_3$ to graphene. As a result, the conduction band has to shift upwards in energy, relative to the graphene Dirac cone, as shown in Fig.~\ref{bands_zoom}d). 
 
Our calculations show that, as we ramp the electric field, the energy bands at the Fermi energy go from a strong hybridization regime in which we expect the bilayer to behave as a half-metal, towards a regime dominated by spin-split graphene Dirac cones, with the flat bands of CrI$_3$ higher in energy and disentangled from those of graphene.  In the second case, both spin channels of graphene will contribute to its conductance, increasing its value by a factor of two. In parallel with the increase in conductance, the spin polarization at the Fermi energy decreases.  

Interestingly, in the non-resonant regime,  the spin splitting of the Dirac cones does depend on the electric field. The effective interlayer exchange is ferromagnetic, as the sign of the spin splitting is the same for the graphene and CrI$_3$ bands. As expected from Eq. (\ref{splitting}), as we increase the field, pulling apart the CrI$_3$ bands from the Dirac point states,  the magnitude of the splitting goes down.  Thus, the splitting can be controlled electrically, and keeping the interface in the non-resonant regime, it can be tuned from 48~meV at zero field, to 28~meV for a field of 100~meV/\AA. The smooth crossover from the hybridization towards the conventional spin proximity effect highlight that both are generated by the same underlying mechanism, namely, the hybridization of the non-magnetic states with spin split states of the ferromagnet.
  

The hybridization proximity effect is by no means specific to CrI$_3$. First, it can be anticipated other magnetic insulators with magnetic ions with a less than half-full  {\it d}-shells in a octahedral environment, 
such as CrBr$_3$, CrCl$_3$, VX$_3$  (X=I, Br, C), CrBrS  also have majority spin  $e_g$ spin-polarized conduction bands. 
In addition,  previous computational work modelling graphene on top of magnetic insulators, such as CrBr$_3$~\cite{behera2019,lyu2022},  CrI$_3$~\cite{zhang2018,cardoso2018}, RuCl$_3$~\cite{mashhadi2019,rizzo2020},  FeCl$_3$~\cite{wei2020} show spin-dependent hybridization of the graphene bands and self-alignment with the spin-split conduction bands of these materials.  

In summary, we propose that  a strong proximity effect that occurs when a magnetic insulator is placed in proximity with graphene. The effect is the consequence of spin-selective  hybridization of the states at the Fermi surface of the conductor with a spin polarized band of the ferromagnet.  The energy of the two types of bands naturally aligns when charge is transferred at the interface to compensate for the work function difference. We claim this will occur naturally when graphene is placed on a variety of  magnetic insulators with spin polarized conduction bands. We also show that the strength of hybridization magnetic proximity can be further tuned by application of an off-plane electric field.  Our DFT calculations of the Van der Waals heterostructure Gr/CrI$_3$ confirm the theoretical scenario.

{\it Acknowledgments}
This work was partially supported by MaX -- MAterials design at the eXascale -- a European Centre of Excellence funded by the European Union's program 
HORIZON-EUROHPC-JU-2021-COE-01 (Grant No. 101093374).
JFR and AC  acknowledge financial support from 
 FCT (Grant No. PTDC/FIS-MAC/2045/2021),
 the Swiss Science National Foundation Sinergia (Grant Pimag),
the European Union (Grant FUNLAYERS
- 101079184).
JFR acknowledges funding from  
FEDER /Junta de Andaluc\'ia, 
(Grant No. P18-FR-4834), 
Generalitat Valenciana funding Prometeo2021/017
and MFA/2022/045,
and funding from MICIIN-Spain (Grant No. PID2019-109539GB-C41).
AHM was supported by the U.S. Department of Energy, Office of Science, Basic Energy Sciences, under Award No.~DE-SC0022106.

\bibliographystyle{apsrev4-1}
\bibliography{refs}

\end{document}